\documentclass[twocolumn,twocolappendix,showpacs,superscriptaddress,groupedaddress,
  nofootinbib,floatfix]{aastex63}

\pdfoutput=1 %for arXiv submission
\usepackage{amsmath,amsfonts,amstext,amssymb}
\usepackage[T1]{fontenc}
\usepackage[utf8]{inputenc}
\usepackage{graphicx}
\usepackage{rotating}
\usepackage{verbatim}
\usepackage{hyperref}
\usepackage{xspace}
\usepackage{enumerate}
\usepackage{xcolor}

\newcommand{\be}{\begin{equation}}
\newcommand{\ee}{\end{equation}}

\newcommand{\AthenaK}{\texttt{AthenaK}\xspace}
\newcommand{\tGRAthena}{\texttt{GR-Athena++}}
\newcommand{\GRAthena}{\tGRAthena\xspace}

\newcommand{\tAthena}{\texttt{Athena++}}
\newcommand{\Athena}{\tAthena\xspace}

\newcommand{\tBAM}{\texttt{BAM}}
\newcommand{\BAM}{\tBAM\xspace}
\newcommand{\tCactus}{\texttt{Cactus}}
\newcommand{\Cactus}{\tCactus\xspace}
\newcommand{\tGRChombo}{\texttt{GRChombo}}
\newcommand{\GRChombo}{\tGRChombo\xspace}

\newcommand{\tDendrogr}{\texttt{Dendro-GR}}
\newcommand{\Dendrogr}{\tDendrogr\xspace}

\newcommand{\tTwoPunctures}{\texttt{TwoPunctures}}
\newcommand{\TwoPunctures}{\tTwoPunctures\xspace}

\def\p{\partial}

\def\e{{\rm e}}
\def\i{{\rm i}}
\def\f{{\rm f,~BH}}

\def\GMc2{{\rm G M_{\odot} c^{-2}}}

\def\B{\mathcal{B}}
\def\I{\mathcal{I}}
\def\M{\mathcal{M}}
\def\O{\mathcal{O}}
\def\R{\mathcal{R}}

\def\l{\ell}

\def\kt2{\kappa^\text{T}_2}

\def\kt2{\kappa^\text{T}_2}

\def\D{\mathrm{D}}

\def\2nd{2^\mathrm{nd}}
\def\4th{4^\mathrm{th}}
\def\6th{6^\mathrm{th}}
\def\8th{8^\mathrm{th}}
\def\sp{\delta x}

\def\z4c{$\mathrm{Z}4\mathrm{c}$}
\def\z4oc{$\mathrm{Z}4(\mathrm{c})$}
\def\z4{$\mathrm{Z}4$}
\def\ccz4{$\mathrm{CCZ}4$}

\newcommand{\Mesh}{\texttt{Mesh}}
\newcommand{\MeshBlock}{\texttt{MeshBlock}}
\newcommand{\MeshBlockPack}{\texttt{MeshBlockPack}}
\newcommand{\AMR}{AMR}

\usepackage{color}
\definecolor{cyan}{rgb}{0,0.9,0.9}
\definecolor{orange}{rgb}{0.9,0.5,0}
\definecolor{magenta}{rgb}{1,0,1}
\definecolor{purple}{rgb}{0.8,0.4,0.8}
\definecolor{gray}{rgb}{0.5,0.5,0.5}

\shorttitle{Numerical Relativity with AthenaK}
\shortauthors{Zhu et al.}

\usepackage{CJK}

\begin{document}
\begin{CJK*}{UTF8}{gbsn}
\title{Performance-Portable Numerical Relativity with AthenaK}

\author[0000-0001-9027-4184]{Hengrui \surname{Zhu} (朱恒锐)}
\affiliation{Department of Physics, Princeton University, Jadwin Hall, Washington Road, New Jersey, 08544, USA}
\affiliation{Princeton Gravity Initiative, Princeton University, Princeton, New Jersey, 08544, USA}

\author[0000-0001-5705-1712]{Jacob \surname{Fields}}
\affiliation{Department of Physics, The Pennsylvania State 
University, University Park, PA 16802}
\affiliation{Institute for Gravitation \& the Cosmos, The
Pennsylvania State University, University Park, PA 16802}

\author[0000-0002-0351-6518]{Francesco \surname{Zappa}}
\affiliation{Theoretisch-Physikalisches Institut, Friedrich-Schiller-Universit{\"a}t Jena, 07743, Jena, Germany}
  
\author[0000-0001-6982-1008]{David \surname{Radice}}\thanks{Alfred P.~Sloan Fellow}
\affiliation{Department of Physics, The Pennsylvania State 
University, University Park, PA 16802}
\affiliation{Department of Astronomy \& Astrophysics, The Pennsylvania State 
University, University Park, PA 16802}
\affiliation{Institute for Gravitation \& the Cosmos, The
Pennsylvania State University, University Park, PA 16802}

\author[0000-0001-5603-1832]{James M.~\surname{Stone}}
\affiliation{School of Natural Sciences, Institute for Advanced Study, 1 Einstein Drive, Princeton, NJ 08540, USA}

\author[0000-0003-3558-7684]{Alireza \surname{Rashti}}
\affiliation{Department of Physics, The Pennsylvania State 
University, University Park, PA 16802}
\affiliation{Institute for Gravitation \& the Cosmos, The
Pennsylvania State University, University Park, PA 16802}

\author[0000-0003-2244-3462]{William \surname{Cook}}
\affiliation{Theoretisch-Physikalisches Institut, Friedrich-Schiller-Universit{\"a}t Jena, 07743, Jena, Germany}

\author[0000-0002-2334-0935]{Sebastiano \surname{Bernuzzi}}
\affiliation{Theoretisch-Physikalisches Institut, Friedrich-Schiller-Universit{\"a}t Jena, 07743, Jena, Germany}

\author[0000-0001-6091-2827]{Boris \surname{Daszuta}}
\affiliation{Theoretisch-Physikalisches Institut, Friedrich-Schiller-Universit{\"a}t Jena, 07743, Jena, Germany}

\correspondingauthor{Hengrui Zhu}
\email{hz0693@princeton.edu}

\begin{abstract}
We present the numerical relativity module within \AthenaK, an open source performance-portable astrophysics code designed for exascale computing applications. 
This module employs the Z4c formulation to solve the Einstein equations.
We demonstrate its accuracy through a series of standard numerical relativity tests, including convergence of the gravitational waveform from binary black hole coalescence. 
Furthermore, we conduct scaling tests on OLCF Frontier and NERSC Perlmutter, where \AthenaK exhibits excellent weak scaling efficiency of 80\% on up to 65,536 \texttt{AMD} MI250X GPUs on Frontier (relative to 4 GPUs) and strong scaling efficiencies of 84\% and 77\% on \texttt{AMD} MI250X and \texttt{NVIDIA} A100 GPUs on Frontier and Perlmutter respectively. 
Additionally, we observe a significant performance boost, with two orders of magnitude speedup ($\gtrsim 200\times$) on a GPU compared to a single CPU core, affirming that \AthenaK is well-suited for exascale computing, thereby expanding the potential for breakthroughs in numerical relativity research.
\end{abstract}


\section{Introduction}
The inherent nonlinearity of the Einstein equations necessitates the use of numerical methods to generate dynamical solutions, a field known as numerical relativity \citep{Baumgarte:1998te, Miller:2000zy, Lehner:2001wq, Baumgarte:2010ndz, Lehner:2014asa, Duez:2018jaf, Hilditch:2024nhf}. This approach has been crucial for modeling gravitational waveforms from binary compact object mergers \citep{Pretorius:2005gq, Baker:2007fb, Campanelli:2005dd, Shibata:1999wm}, particularly in light of the increasing detections from the LIGO, Virgo, and KAGRA collaborations (LVK) \citep{Abbott:2016blz, TheLIGOScientific:2016wfe, TheLIGOScientific:2017qsa, akutsu2020overviewkagracalibration}. With LVK's ongoing sensitivity enhancements \citep{Abbott_2020} and the upcoming next-generation detectors \citep{amaroseoane2017laserinterferometerspaceantenna, Punturo:2010zz, Evans:2016mbw, Reitze:2019iox}, there is a pressing need for more accurate gravitational waveform catalogs that cover larger parameter spaces, including eccentric, precessing, and high-mass-ratio black hole binaries \citep{Lousto:2010qx, Nakano:2011pb, lovelace:2021}.

One of the central challenges in numerical relativity (NR) is resolving the wide range of spatial and temporal scales present in simulations. To handle this, NR codes often employ adaptive mesh refinement (AMR). 
A widely used implementation is the {\tt Carpet} thorn \cite{Schnetter:2003rb} on top of the \Cactus{} framework \cite{Goodale:2002a}, which leverages the Berger-Oliver method \citep{Berger:1984zza, Berger:1989a} where nested patch patches of grid with increasing resolution are used to resolve the binary constituents and track their evolution in time. 
NR codes building on top of this framework include {\tt Llama} \citep{Pollney:2009yz,Reisswig:2012nc}, {\tt McLachlan} \citep{brown2009turduckeningblackholes}, and {\tt LEAN} \citep{sperhake2007binaryblackholeevolutions}. Other NR codes, including \BAM{} \citep{Brugmann:2008zz}, {\tt AMSS-NCKU} \citep{cao2008reinvestigationmovingpunctured}, and \GRChombo{} \citep{Clough:2015sqa}, also use the Berger-Oliger AMR approach. 

This approach, however, incurs heavy performance penalties due to the overhead for synchronization of data between patches at different resolution, thereby decreasing the scaling performance on modern high-performance computing (HPC) systems \citep{stout1997adaptiveblockshigh}. 
This leads to the adoption of block-based oct-tree AMR in codes like \Dendrogr \cite{Fernando:2018mov,Fernando:2022php} and \GRAthena \citep{Daszuta:2021ecf,Cook:2023bag,Daszuta:2024chz,Daszuta:2024ucu}. 
Both of the codes have demonstrate excellent scaling on large CPU clusters. 

All of the aforementioned codes uses finite differencing with Cartesian coordinate. Other methods include pseudo-spectral techniques, exemplified by {\tt SpEC} \citep{Szilagyi:2009qz}, where the domain is decomposed into patches of topological spheres and cylinders. 
Another example is the {\tt bamps} code, which uses pseudo-spectral method for spacetime variables but Discontinuous Galerkin (DG) for matter \citep{Hilditch:2015aba,Bugner:2015gqa}.
A novel hybrid finite difference and DG method is used in {\tt SpECTRE} \citep{Kidder:2016hev,spectrecode} and \texttt{Nmesh} \citep{Tichy:2022hpa}.
A more recent innovation aims to eliminate the need for AMR by generating problem-specific curvilinear grids, as seen in {\tt SENR/NRPy+} \citep{Mewes:2020vic, mewes2018numericalrelativityspherical, ruchlin2018senrnrpynumericala}.

%Despite the success of many existing codes, the increasing demand for higher resolution and faster turnaround times, coupled with the advent of exascale computing, has exposed limitations in scalability and GPU utilization in these traditional NR codes.
The emergence of exascale computing platforms provides unprecedented amounts of computational power, but to fully harness this, numerical relativity codes must efficiently run on GPU architectures and scale well across thousands of nodes. 
Most existing codes are limited by either their reliance on CPU-based architectures or exhibit poor scaling on large clusters. 
As a result, next-generation NR codes capable of operating on exascale systems are essential to further scientific progress.

To address these challenges, we introduce \AthenaK, a performance-portable extension of the \Athena astrophysics code \citep{white2016extensionathenacode, felker2018fourthorderaccuratefinite, stone2020athenamathplusmathplus}, developed using the \texttt{Kokkos} library \citep{kokkos}. \texttt{Kokkos} abstracts machine-specific dependencies, allowing \AthenaK to fully utilize the computational power of exascale clusters for a broad range of astrophysical simulations.

In this paper, we describe the implementation of the vacuum Einstein equations in Z4c formulation within \AthenaK, using oct-tree AMR.
We demonstrate the accuracy of the code through a series of tests, including the convergence of gravitational waveforms from binary black hole coalescence.
Additionally, we introduce new AMR criteria that enable the simulation of binary black holes with large mass ratios.
Strong and weak scaling results from the \texttt{Frontier} supercomputers show \AthenaK achieves 80\% weak scaling efficiency on 65,536 GPUs, providing a clear path towards exascale numerical relativity. This work is part of a three-paper series describing \AthenaK's key functionalities \citep{Stone.6.24, Fields.6.24}, available on GitHub\footnote{The code is available at \href{https://github.com/IAS-Astrophysics/athenak}{https://github.com/IAS-Astrophysics/athenak}}.

\AthenaK is not the only, nor the first, effort to leverage GPUs for numerical relativity applications.
For instance, \texttt{AsterX} and \texttt{GRaM-X} are notable examples \cite{Kalinani:2024rbk,Shankar:2022ful}.
Both of these codes solve the GRMHD equations coupled with the Einstein equations in the Z4c formulation, using the \texttt{CarpetX} \AMR{} driver \cite{Schnetter:2003rb} within the \texttt{Einstein Toolkit} \cite{EinsteinToolkit:2024_05}, which is built on \texttt{AMReX} \cite{Zhang2019,doi:10.1177/10943420211022811}.
However, the convergence and waveform accuracy of the Z4c solvers are yet to be demonstrated. 
Additionally, \Dendrogr offers a GPU extension as presented in \cite{Fernando:2022zbc}, but scaling results have been limited to 16 GPUs. 
The code is primarily written in \texttt{CUDA}, which restricts its compatibility to NVIDIA systems, although there is preliminary support for AMD GPUs.
An similar effort to \AthenaK is ongoing to develop a numerical relativity module in the \texttt{Phoebus} code \citep{phoebus}, which is based on the \texttt{Parthenon} framework \citep{Grete_2022}.

In this paper, we show that \AthenaK is performance portable, shows better scaling, and gives accurate and convergent waveforms.
The paper is organized as the following: 
In section~\ref{sec:z4c}, we recap the choice of Z4c formulation for solving the Einstein equations.
In section~\ref{sec:num}, we describe the numerical implementation for vacuum GR, focusing on algorithmic \textit{differences} with \GRAthena. 
We present a series of test problems, including convergence of binary black hole waveforms, in section~\ref{sec:test}. 
All tests were run on GPUs.
We then present performance and scaling results in section~\ref{sec:scale}. 
Lastly, we conclude in section~\ref{sec:conclude}. 

\section{z4c formulation}\label{sec:z4c}
As in \GRAthena, we evolve the Einstein equation in the conformally-decomposed Z4 formulation \citep{Bona:2003fj}, or Z4c \citep{Bernuzzi:2009ex,Hilditch:2012fp}.
The continuum form of our equations and gauge choices are exactly the same as in \GRAthena{} \citep{Daszuta:2021ecf}, to which we refer interested readers for details. 
Here, we briefly summarize advantages of the Z4c formulation. 

Our choice of the Z4c formulation for solving the Einstein field equations is driven by its ability to integrate the strengths of two prominent approaches \citep{Gundlach:2005eh}: 
the BSSNOK formulation \citep{Nakamura:1987zz,Shibata:1995we,Baumgarte:1998te} and the generalized harmonic gauge (GHG) formulation \citep{Friedrich:1985,Pretorius:2005gq,Lindblom:2005qh}.
First, like the GHG formulation, Z4c removes the zero-speed characteristic in the constraint subsystem, which could cause constraint violation to build up especially in the presence of matter; furthermore, it admits a natural constraint damping scheme \citep{Cao:2011fu,Weyhausen:2011cg,Hilditch:2012fp}.
Second, like in BSSNOK, Z4c is compatible with the puncture scheme for evolving black holes, where coordinate singularities are explicitly advected in the computational domain.
The puncture scheme avoids the need for excision, where the region interior to the apparent horizons are removed. 
Due to the non-locality of the horizon finding routine, avoiding excision would greatly improves parallel efficiency. 

We refer readers to Eqn. 8-13 of \cite{Daszuta:2021ecf} for the exact form of equations used in our implementation. In addition to these equations, one must also specify the gauge conditions to close the system. Same as in \GRAthena, we use the Bona-M\'asso lapse \citep{Bona:1994a} and the gamma-driver shift \citep{Alcubierre:2003hr} (Eqn. 22 in \cite{Daszuta:2021ecf}). 

\section{Numerical Implementation}\label{sec:num}
We refer interested readers to \cite{Stone.6.24} and \cite{stone2020athenamathplusmathplus} for details regarding the numerical infrastructure for \AthenaK. 
Here we briefly recap key numerical features in \AthenaK{} for evolving the vacuum Einstein equations, highlighting the difference in design choice with its predecessor \GRAthena{}. 

As in \Athena, the computational domain over which a set of physics are evolved is abstracted in a class named \Mesh{}. 
The \Mesh{} contains rectangular blocks known as the \MeshBlock s. 
Different from \Athena, the \MeshBlock s are further organized into \MeshBlockPack s to reduce the number of kernel launches, which greatly improves performance on GPUs \citep{Grete_2022}. 

Each \MeshBlock{} consists of an active region and a ghost region. Data in the ghost region must be communicated between \MeshBlock s and is used to set boundary conditions.
At each stage of the time integration process, where we employ a family of explicit Runge-Kutta methods, the ghost zones are filled by either the active region of neighboring \MeshBlock s or the boundary conditions.
When adaptive or static mesh refinement (\AMR{} or SMR respectively) is used, neighboring \MeshBlock s may have different resolutions, restricted to factors of two in \AthenaK. On these refinement boundaries, a set of prolongation and restriction operations are performed to fill the ghost cells with the required accuracy.
Since we use higher-order finite differencing to evolve the Z4c equations—unlike the fluid evolution which uses second-order finite volume approach—the boundary communication strategy at refinement boundaries must be modified. We detail these modifications below.

\subsection{Modifications to refinement boundary communication}

\begin{figure}[tt]
  \includegraphics[width=0.46\textwidth]{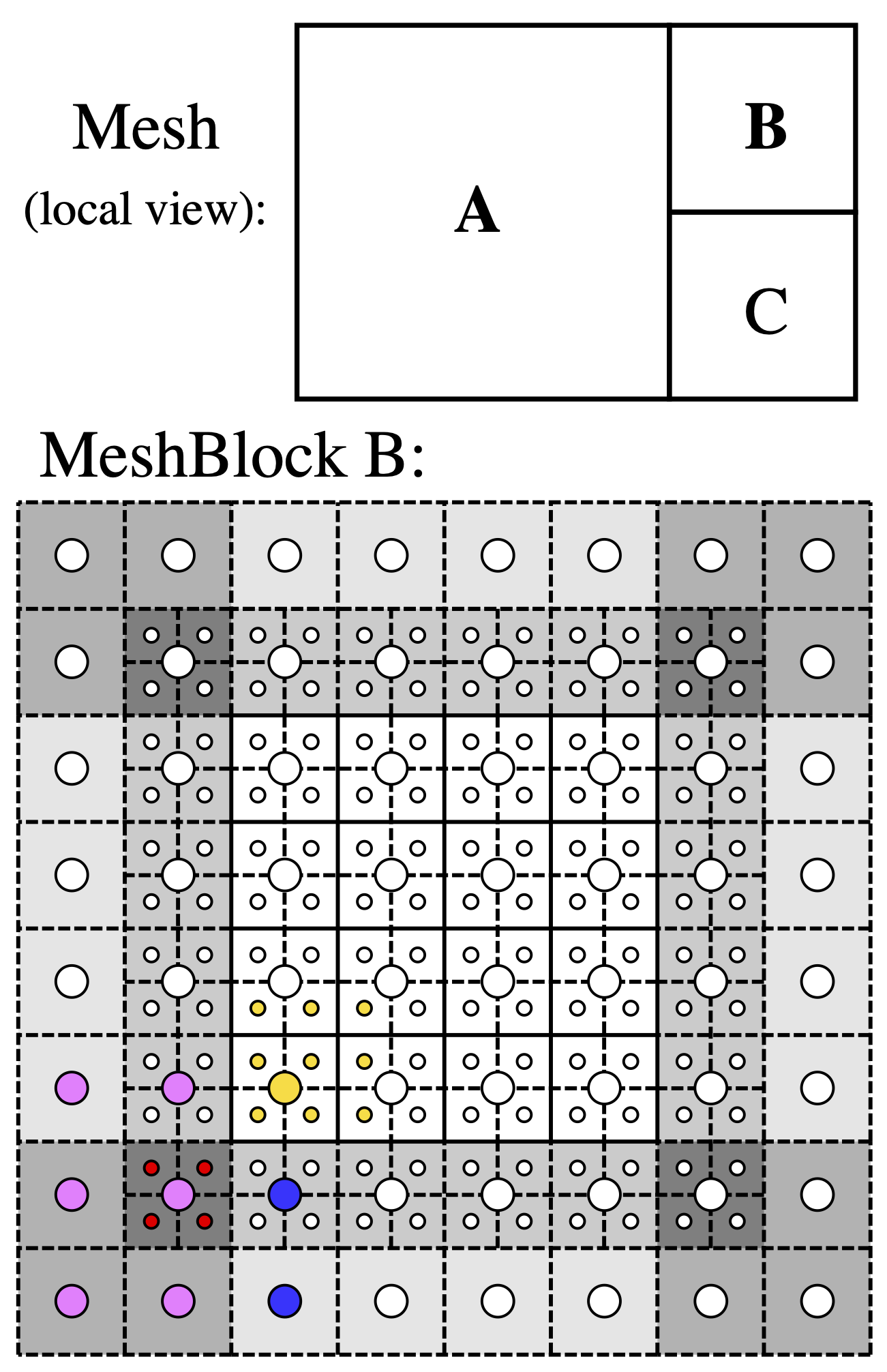}
  \caption{Top: schematic for \MeshBlock s at refinement boundary. The squares show the \textit{active} region of each \MeshBlock. Bottom: nodal structure inside \MeshBlock{} B. The smaller circles represent the nodes in the \MeshBlock, and larger circles representing the coarse buffers used in \AMR{} communications. The active region is unshaded, and the ghost region is shaded with different grey levels. We also draw schematics for the prolongation and restriction stencils. The smaller yellow circles show the restriction stencil to fill the coarse buffer node highlighted in yellow, and all colored large circles make up the prolongation stencil to fill in the red little circles in the fine ghost zone.  }
  \label{fig:p_and_r}
\end{figure}

As in the fluid sector, the boundary communication at mesh refinement boundary is done with a set of prolongation and restriction operators. 
The top panel of Fig.~\ref{fig:p_and_r} illustrates three \MeshBlock s at a refinement boundary. 

From the perspective of \MeshBlock~A, the communication is the same to that for the fluid sector: 
\begin{enumerate}
    \item Pack and send active nodes that overlaps with the coarse buffer of nearby finer \MeshBlock s, for example, the large pink dots in Fig.~\ref{fig:p_and_r}. 
    \item Receive nodes from active region of coarse buffers of neighboring finer \MeshBlock s, e.g. the large yellow dot in Fig.~\ref{fig:p_and_r}. 
\end{enumerate}
See Fig. 4 and discussion in Section \textit{2.1.3} of \cite{stone2020athenamathplusmathplus} for detailed illustrations. 

The perspective for \MeshBlock~B is slightly modified:
\begin{enumerate}
    \item Restrict the data in the active region to fill the coarse buffer. 
    \item Send coarse buffer to neighboring \MeshBlock s, to fill either their ghost zone (for neighbors with one less level of refinement, i.e. \MeshBlock~A) or the ghost region for their coarse buffer (for neighbors with the same refinement level, i.e. \MeshBlock~C)
    \item Send ghost zone to neighboring \MeshBlock s at the same level, i.e. \MeshBlock~C. 
    \item Receive coarse buffer and ghost zones from neighbors to fill the ghost zone and parts of the ghost zone for the coarse buffers that overlaps with the prolongation stencils. 
    \item Prolongate the coarse buffer to fill the fine ghost zones. 
\end{enumerate}

The main distinction between the above procedure with that for the fluid, as described in \cite{stone2020athenamathplusmathplus}, is that the coarse buffers must be passed between neighboring \MeshBlock s at the same level (namely B and C), as the prolongation stencil now overlaps with these nodes. We now discuss the change in the restriction and prolongation operators.

\subsubsection{Restriction}
For evolving the fluid with finite volume, each nodes store the cell average, so the restriction operation becomes a trivial averaging. 
This is no longer the case with finite differencing. 
To fill the coarse buffer for the fine region, we use high-order Lagrange polynomial interpolation, which calculates the value for the desired node with coordinate $x$ as a weighted sum, given a set of $\{(x_i, f(x_i))\}$ in the interpolation stencil. 
In one dimension, the weights are given by:
\begin{equation}
\ell_i(x) = \prod_{\begin{smallmatrix}0\le m\le k\\ m\neq j\end{smallmatrix}} \frac{x-x_m}{x_i-x_m}
\end{equation}
Then, the value at x is given by:
\begin{equation}
    f(x) = \sum_{0\leq i \leq k} l_i(x) f(x_i)
\end{equation}
This procedure can be generalized to two and three spatial dimensions by taking the outer-product between the interpolation weights for each dimension. 

The order of convergence for Lagrange interpolation is $n+1$, where $n$ is the number of points in the stencil for each spatial dimension. 
In \AthenaK, we choose the interpolation stencil size to be $n = N_g+1$, where $N_g$ is the number of ghost cell, so that the error term converges away at $3^{\mathrm{rd}}$ order for 2 ghost, and $5^{\mathrm{th}}$ order for 4 ghost.\footnote{By default, we use $2^{\mathrm{nd}}$ order spatial differencing when using 2 ghost cells, and $6^{\mathrm{th}}$ order with 4 ghost, while the time integrator is kept at $4^{\mathrm{th}}$ order. In practice we never find the prolongation and restriction operator to dominate over the spatial differencing error. }

In the lower panel for Fig.~\ref{fig:p_and_r}, we show the restriction stencil for two ghost cells. To fill the coarse buffer in the corner of the active region (unshaded), the large yellow dot as an example, we use the stencil consisting of the 9 small yellow dots (this becomes 27 nodes in three spatial dimension). 
We note that the stencil is always within the active region. Such a choice is made to improve locality for the restriction step. 
Albeit the asymmetry in the interpolation stencil, we have yet to find any sacrifice in accuracy. 

\subsubsection{Prolongation}
The prolongation operator also uses Lagrange interpolation polynomial. 
In Fig.~\ref{fig:p_and_r}, we highlight the four fine ghost cells (little red dots) in the corner of \MeshBlock{} B (or edges in three dimension) that needs to be filled during boundary communication. The prolongation stencil consists of the 9 colored large dots: yellow one filled from restriction in the active region of \MeshBlock{} B, blue ones from the coarse buffer of \MeshBlock{} C, and pink ones from fine nodes in \MeshBlock{} A. 

\subsection{Cell-Centered vs Vertex-Centered Finite Differencing. }
The aforementioned \AMR{} strategies are also different from those in \GRAthena, because here we use a Cell-Centered (CC) finite differencing scheme as oppose to Vertex-Centered (VC). \footnote{A recent CC extension, referred to as CX in \cite{Daszuta:2024ucu}, is added to \GRAthena. We defer a comparison between the two schemes to a future work.}
In a VC scheme, the finite differencing nodes are located at cell vertices, so that the coarse nodes overlap with some fine nodes, making the restriction step trivial. 
However, this comes with some complications when coupling the Einstein equation with fluid equations which are typically solved using finite volume methods: one must interpolate the metric components and the gauge variables to the cell center, which serves as effective source terms for the fluid equations, and the stress-energy tensor for the fluid back to the cell vertices for evolving the spacetime. 
These extra interpolation operations add to the total computational cost when evolving spacetime with matter \citep{Daszuta:2024ucu}. 
In our CC scheme, the fluid variables are stored at the same location as the Z4c variables, making these interpolation operations unnecessary.

\subsection{New \AMR{} criteria}
In addition to the \AMR{} algorithms implemented in \GRAthena, see, e.g. \cite{Rashti:2023wfe}, we implement two additional \AMR{} criteria that are agnostic to the puncture location, based on the conformal factor $\chi$. 
The conformal factor $\chi$ goes to zero towards the punctures, naturally tracking the location of the black holes. 
Since it is crucial to adequately resolve the horizon to avoid constraint violation propagating outwards, we require more resolution near the punctures. 
Therefore, $\chi$ could be used as a good proxy for specifying region of the computational domain to refine for binary black hole problems. 
Here, we present two $\chi$-based \AMR{} criteria. 

\subsubsection{$\chi-\textrm{min}$ criterion}

For the $\chi$-min criterion, we assess whether the value of $\chi$ within each \MeshBlock{} falls below a specified threshold. If it does, the \MeshBlock{} is refined, provided it has not already reached the maximum level of mesh refinement. The \AthenaK \AMR{} framework maintains a two-to-one refinement condition between neighboring \MeshBlock s, allowing the resolution to gradually decrease as one moves away from the vicinity of the punctures.
In the BAM calibration binary black hole run, we observed that this criterion typically results in fewer \MeshBlock s being created or destroyed compared to the commonly used puncture-tracker-based $L^2$ criterion.
A snapshot of the refinement structure in the strong field for the calibration run is shown in Fig.~\ref{fig:snapshot}. 
A detailed comparison with the $L^2$ criterion is deferred to a future work.

\subsubsection{$d\chi-\textrm{max}$ criterion}

\begin{figure}[t]
  \includegraphics[width=0.45\textwidth]{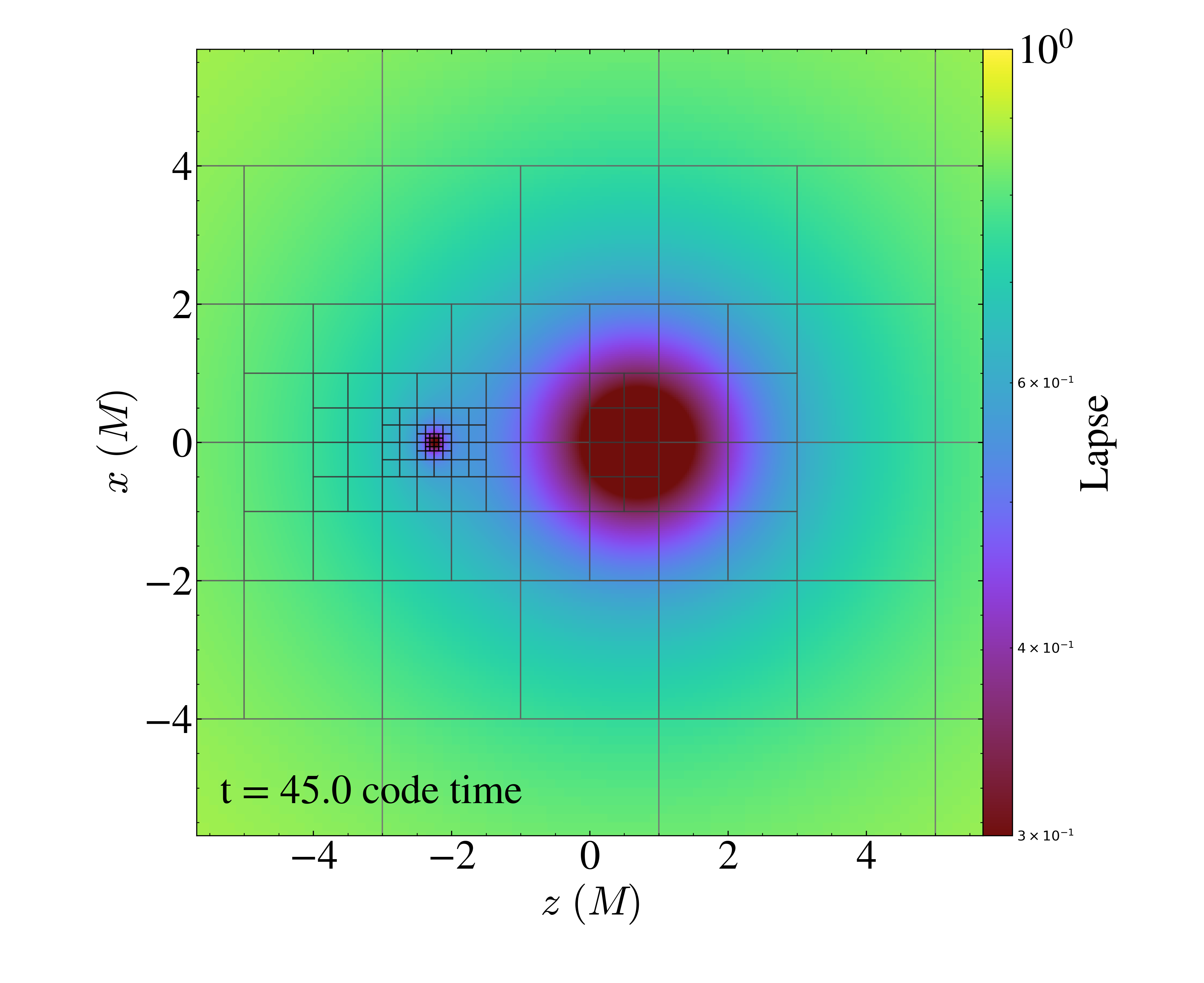}
  \caption{Mesh refinement structure of a black hole binary with a mass ratio of 20 using the $d\chi-\textrm{max}$ refinement criteria. The smaller hole automatically receives 4 more levels of mesh refinement due to its larger gradient in the conformal factor. }
  \label{fig:dchi_amr}
\end{figure}

The $\chi$ criterion works well for equal mass ratios, but in asymmetric binaries, it tends to over-resolve the interior of the larger black hole. 
To address this, we also implement a $d\chi$-max criterion. 
At each refinement step, we check whether the gradient of $\chi$ exceeds a certain threshold within a \MeshBlock. 
If it does, the \MeshBlock{} is refined, unless it has already reached the maximum refinement level.
Since this criterion is based on the gradient of the conformal factor rather than $\chi$ itself, it naturally results in higher levels of refinement around the smaller black hole. 
We illustrate the resulting grid structure during a head-on collision with a reference mass ratio of 20 in Fig.~\ref{fig:dchi_amr}. 
The $d\chi$ criterion effectively tracks the black holes and naturally produces four additional levels of mesh refinement near the smaller puncture, avoiding over-resolution of the larger black hole's interior.

\subsubsection{derefinement at the wave zone}
A common issue for both of the \AMR{} criteria above is that the resolution could fall off fast away from the black holes.
This is problematic since the gravitational waveform is usually extracted at large radii ($\gtrsim 100~r_g$), and a low resolution at the wave zone could significantly impact its accuracy.
To address this, we implement an additional \textit{radius} criterion, where all meshblocks within certain radius of a given point are kept above a specified refinement level. 

\subsection{Other Diagnostics}
As in \GRAthena, we implement two key diagnostics for binary black hole evolution: puncture trackers and wave extraction. The puncture trackers are initialized at the puncture location in the initial data and advected across the domain by integrating the shift vector at each time step. To extract gravitational waves, we compute the outgoing Weyl scalar $\Psi_4$ using a coordinate tetrad, then interpolate $\Psi_4$ onto geodesic spheres at various radii and decompose it into spin-weighted spherical harmonics. Ongoing work includes a horizon finder and outputting worldtube data for Cauchy Characteristic Extraction \citep{Bishop:1996gt,Reisswig:2009rx,Bishop:2016lgv,Barkett:2019uae,Moxon:2020gha,Moxon:2021gbv}.

\section{Code Tests}\label{sec:test}

To test the correctness and accuracy of our implementation, we perform a series of tests and compare the results to \BAM{} and \GRAthena in this section. 
All tests in this section are done on \texttt{NVIDIA} A100 GPUs. 

\subsection{Linear wave convergence}
To test the robustness of our time integrator as well as finite differencing scheme, we perform a linear wave convergence test in full three spatial dimension \citep{Alcubierre:2003pc}.
We set periodic boundary condition, and propagate the wave along the diagonal of the \Mesh, with an amplitude of $10^{-8}$.
We then calculate the $L^1_{\rm RMS}$ error after one cycle of propagation at time $T$, given by:
\begin{equation}
    L^1_{\rm RMS} = \sqrt{\sum_{a,b\in\{x,y,z\}}\left(\int_{\rm mesh} dv |g_{ab}(T)-g_{ab}(0)|/ \textrm{Vol.}\right)^2}~,
\end{equation}
where $\textrm{Vol.}$ is the volume of the \Mesh. 

\begin{figure}[tt!]
  \includegraphics[width=0.45\textwidth]{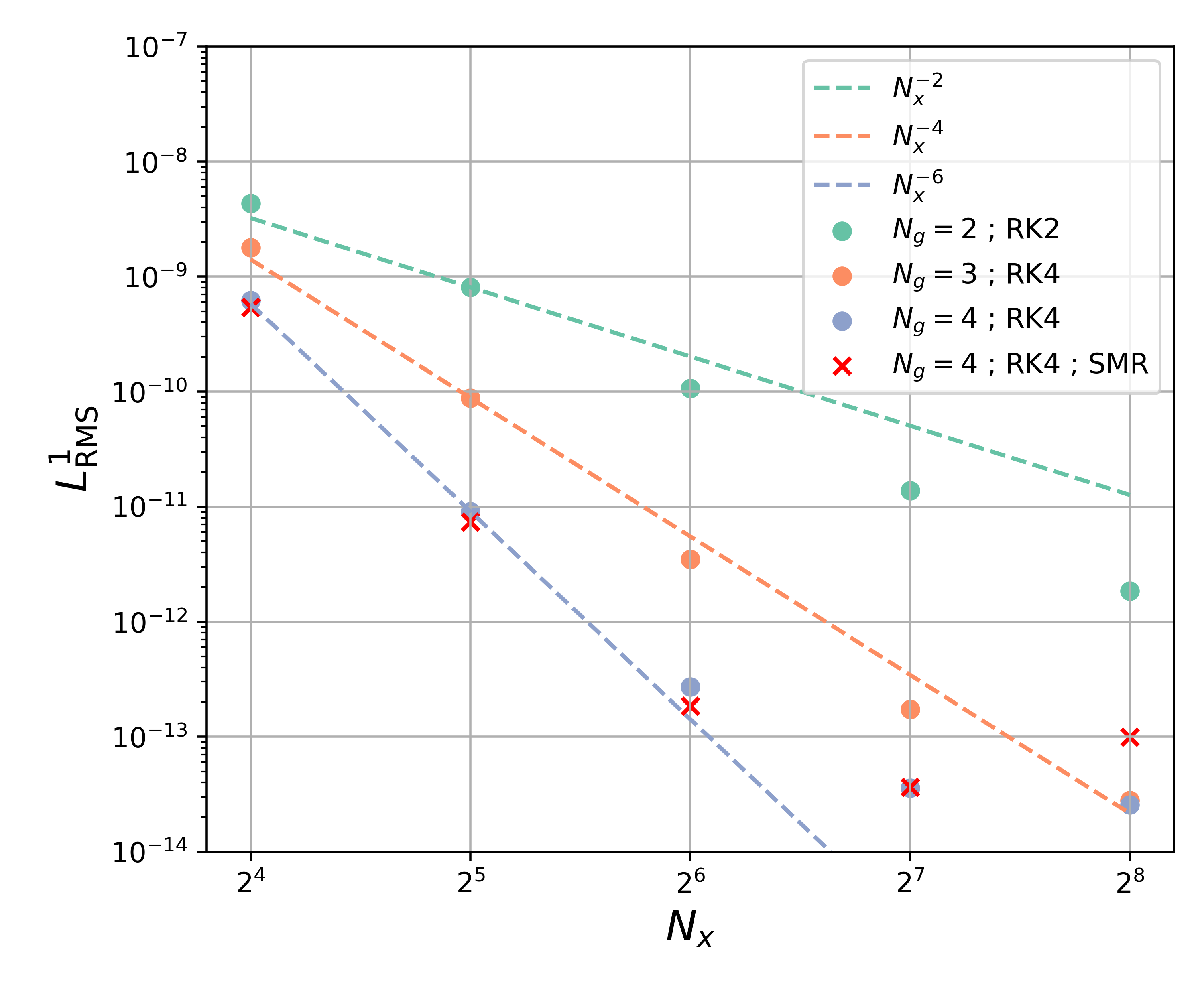}
  \caption{$L^1_{\rm RMS}$ error of linear gravitational wave after one cycle of propagation using different numerical schemes. In particular, the red crosses are from runs with a quadrant of the \Mesh{} refined, to test the accuracy for the prolongation and restriction operators described in Sec.~\ref{sec:num}. }
  \label{fig:wave}
\end{figure}

We try the evolution with three different combination of spatial differencing orders and time integrator. As shown in Fig.~\ref{fig:wave}, we find all the algorithms converges at the expected order. For the combination of $N_g=4$ and RK4, we find it to be converging at $6^{\rm th}$ order, indicating that in this regime the spatial differencing error is dominating over that of the time integrator, thereby the convergence order of the scheme is higher than that of the time integrator.

To test the accuracy for the prolongation and restriction operators, we perform the same linear wave evolution but now with a corner of the \Mesh{} refined at one level higher. We plot the results as red crosses in Fig.~\ref{fig:wave}. We find that the RMS error is consistently lower than the same run without the mesh refinement in the corner, except for the last data point, where the added number of operations causes accumulation of round-off error and therefore an overall increase in the RMS error.

\begin{figure}[tt!]
  \includegraphics[width=0.45\textwidth]{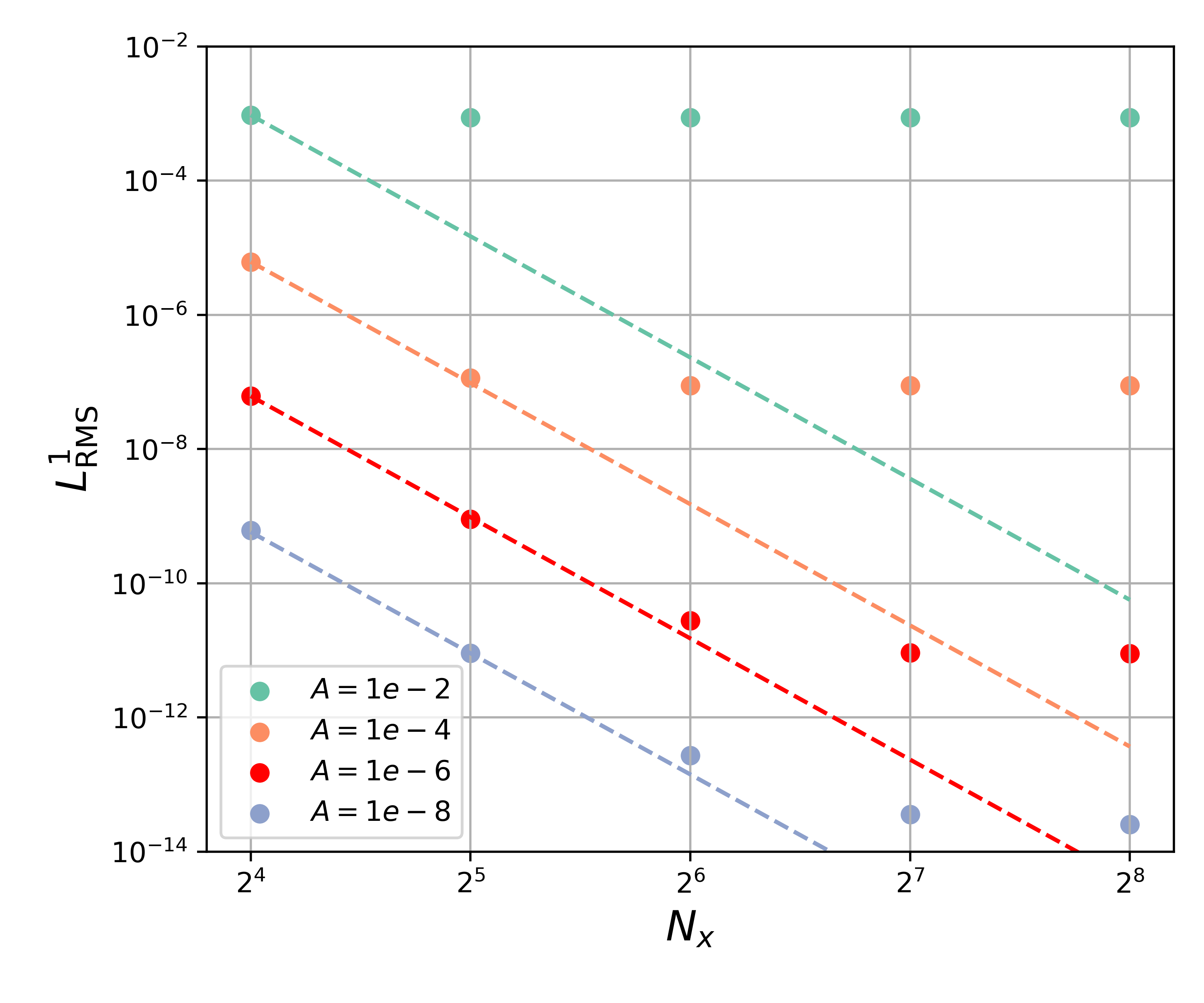}
  \caption{Linear wave tests for different initial amplitude reveal a nonlinear effect. The dashed lines show the expected rate of convergence. Note that the plateau in the RMS error becomes quadratically spaced in the wave amplitude, indicating that they arise from quadratic effects. }
  \label{fig:wave_nl}
\end{figure}

At high resolution (lower-right corner of Fig~\ref{fig:wave}), we find that the RMS error plateau. 
This is due to nonlinear effects (coming from the fact that the solution of the EFE is not the same as the advection equation).
To further explore this, we run the same linear wave test with $6^{\rm{th}}$ order spatial differencing and RK4 but with different wave amplitude. The results are shown in Fig.~\ref{fig:wave_nl}, where we find that the plateau in the RMS error becomes quadratically spaced in the wave amplitude, indicating the cause of this error is indeed a quadratic effect, with a coefficient of order 10.

\begin{figure}[t]
  \includegraphics[width=0.45\textwidth]{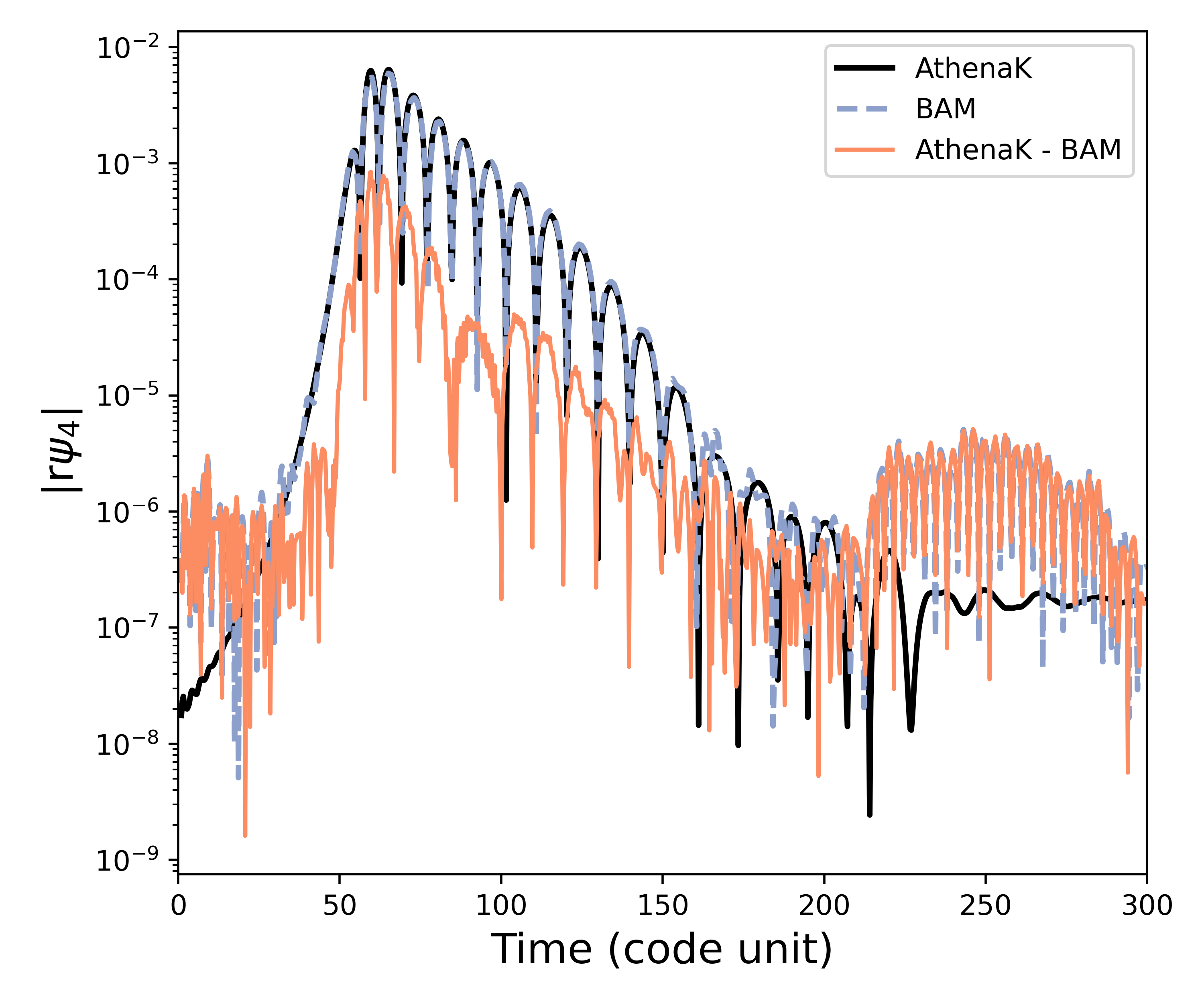}
  \caption{\AthenaK \textit{vs.} \BAM waveform for a single spinning puncture. Here, we plot the $l=2$, $m=0$ component of the Weyl scalar extracted at a coordinate radius of 50 M. The difference between the two code remains small during the course of the evolution. }
  \label{fig:one_punc}
\end{figure}

\subsection{Single Puncture Test}
Before simulating a full binary black hole coalescence, we first demonstrate the robustness of \AthenaK{} for evolving a single spinning puncture.
Using initial data generated by the \TwoPunctures library, as employed in similar tests of the \BAM code in \cite{Hilditch:2012fp} and \GRAthena code in \cite{Daszuta:2021ecf}, we simulate the evolution of a single spinning puncture, representing a rotating black hole with a dimensionless spin parameter $a=0.5$. We initialize two black holes: one with the target mass of $1M$ and spin $a=0.5$, and another with a negligible mass of $10^{-12}M$ and zero spin, separated initially by $10^{-5}M$. 
The smaller black hole is not numerically resolved by the evolution code and only sources a small perturbation. 
Therefore this initial data can be treated as the target rotating black hole with some small perturbation. 
We utilize the static mesh refinement to construct a grid around the puncture, extending to $\pm 1024 M$ in each spatial dimension. The resolution is set to be consistent with the \BAM evolution both at the puncture ($\sp = 0.08333 M$) and in the wave zone ($\sp = 0.66667 M$).
As shown in Fig.~\ref{fig:one_punc}, the $l=2$, $m=0$ harmonic of the gravitational waveform from \AthenaK using $6^{\rm th}$ order spatial differencing is consistent with that from \BAM, which used $4^{\rm th}$ order finite differencing. 

\subsection{Binary Black Hole Evolution}\label{sec:bbh}
Lastly, we test the correctness and convergence of the gravitational waveform from binary black hole coalescence. 
For this we use the \BAM{} calibration binary black hole problem \cite{Bruegmann:2006ulg}, which consists of an equal-mass, non-spinning quasi-circular binary that merges in three orbits with \TwoPunctures initial data \cite{Ansorg:2004ds}. 
Here, we compare the waveform from \AthenaK{} with \GRAthena. %\wc{I think here you should make clear it's the calibration BBH problem described in that BAM paper, but you aren't using BAM data, instead \GRAthena data}

Like in \GRAthena \cite{Daszuta:2021ecf}, we choose a K.O. dissipation of 0.5, constraint damping parameters $\kappa_1=0.02$, $\kappa_2 = 0$, and CFL number of 0.25. We use $6^{\rm th}$ order spatial differencing and the RK4 time integrator.

\begin{figure}[t]
  \includegraphics[width=0.45\textwidth]{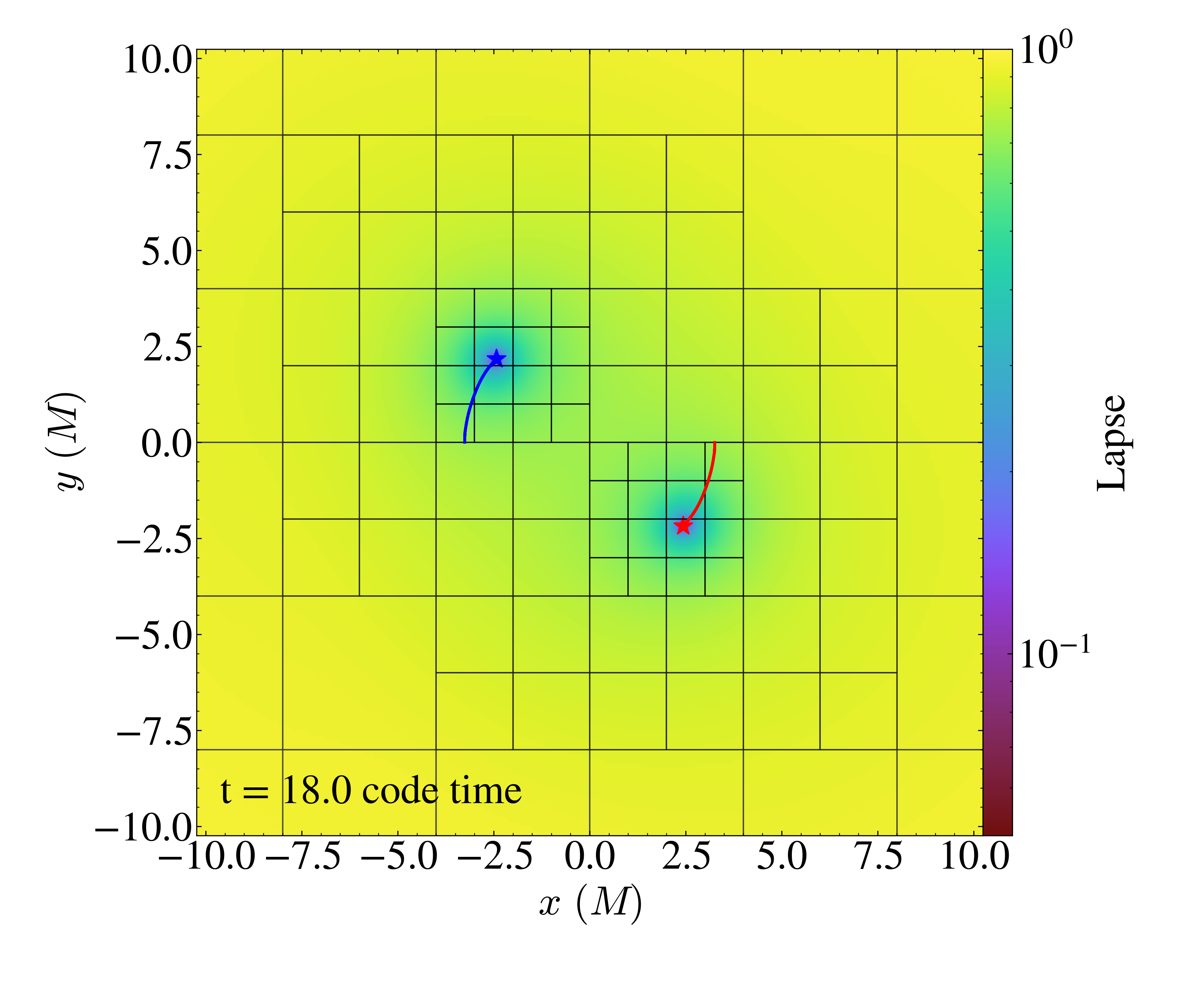}
  \caption{Mesh-refinement structure during the calibration binary black hole run using the $\chi$ criterion. The history of punctures' location (through puncture trackers) are overplotted as blue and red curve respectively. }
  \label{fig:snapshot}
\end{figure}

\subsubsection{Self-Convergence}
We first test the self-convergence of the gravitational waveform, measured with the Weyl scalar $\Psi_4$.
We use a \Mesh{} extending from $-1024$ to $1024$ M in all three spatial dimension, with M measured in the ADM mass of the initial data. 
The root grid consists of $4^3$ \MeshBlock s. 
To resolve the horizons, we use 10 levels of mesh refinement with the $\chi-\textrm{min}$ refinement criteria, with $\chi_\textrm{min}$ threshold set to 0.3. 

We run the calibration runs at three different resolutions.
To demonstrate convergence with \AMR{} enabled, we uniformly increase the resolution by factors of two by changing the number of points in each \MeshBlock{} ($32^3$, $64^3$, and $128^3$) but without changing the overall structure of the mesh, so that the structure of the computational domain is maximally consistent between runs with different resolutions. 
The resulting resolutions at the puncture are 0.03125, 0.015625, and 0.0078125 respectively. 

%\wc{I would explicitly give the grid parameters here, especially things like where the outer boundary is, and what the resolution at the punctures is, and which of the above described AMR criteria you use.}

The numerical waveform consists of the continuum limit $\psi_{\mathrm{cont}}(t)$ and an error term that scales polynomially with resolution $\Delta x$, namely:
\begin{equation}
    \psi_{\mathrm{NR}} (t,\Delta x)  = \psi_{\mathrm{cont}}(t) + \xi (\Delta x) ^ n~,
\end{equation}
where $n$ is the convergence order for the algorithm, and $\xi$ is an error coefficient dependent on discretization schemes as well as the physical problem. When comparing waveforms at different resolutions, the residual $\epsilon$ between the two runs is then given by:
\begin{align}
    \epsilon(\Delta x_1,\Delta x_2) &= \psi (t,\Delta x_1) - \psi (t,\Delta x_2) \nonumber \\
    &=\xi ((\Delta x_1)^n - (\Delta x_2)^n)~.
\end{align}
To show order of convergence, we then show that
\begin{equation}
    \frac{\epsilon(\Delta x_1,\Delta x_2)}{\epsilon(\Delta x_2,\Delta x_3)} \simeq \frac{(\Delta x_1)^n - (\Delta x_2)^n}{(\Delta x_2)^n - (\Delta x_3)^n} =: Q_n~, \label{eq:qfactor}
\end{equation}
for appropriate order n. 

\begin{figure}[t]
  \includegraphics[width=0.45\textwidth]{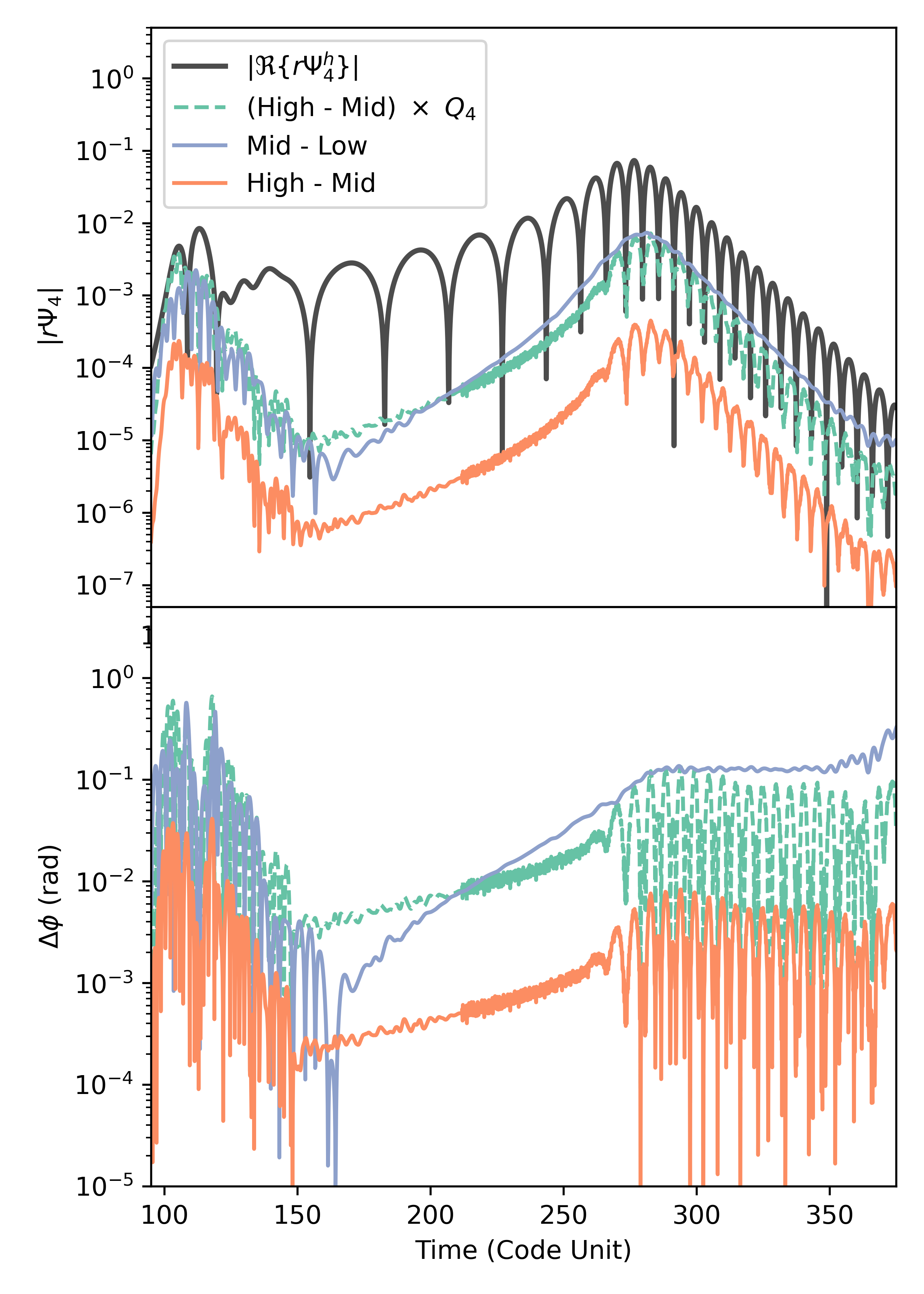}
  \caption{Convergence of gravitational waveform for the calibration binary black hole run. \texttt{Top:} We show the real part of the quadrupolar (${\ell=m=2}$) component of $\Psi_4$ from the high resolution run, and in colored lines we show the residual between mid and low resolution $|\Psi_4^{\rm mid}-\Psi_4^{\rm low}|$ (in purple), and high and mid resolution $|\Psi_4^{\rm mid}-\Psi_4^{\rm low}|$ (in orange). To demonstrate $4^{\rm th}$ order convergence, we scale up the residual between the high and mid resolution by the Q factor defined in Eq.~\ref{eq:qfactor} and plot it as the dashed cyan curve. When compared with the purple curve, clear $4^{\rm th}$ order convergence is seen. \texttt{Bottom:} The same but for the phase difference between the waveforms. }
  \label{fig:conv}
\end{figure}

In Fig.~\ref{fig:conv}, we plot the waveform as well as the residual between different resolutions in terms of the Weyl scalar $\Psi_4$, extracted at a coordinate radius of 100 M. 
The residuals for both the waveform amplitude and the phase are plotted in logarithmic scale to better demonstrate convergence. 
The residual between the high and mid resolution when scaled up by the Q factor is consistent with the that between the low and mid resolution, suggesting a $4^{\rm th}$ order convergence. \footnote{This order of convergence could in principle be improved by adopting techniques developed in \cite{Etienne:2024ncu}, which showed that the numerical error in puncture evolution is dominated by the sharp lapse features at the start of the evolution. We will investigate this in a future work. }

\subsubsection{Comparison with \GRAthena}

\begin{figure}[t]
  \includegraphics[width=0.45\textwidth]{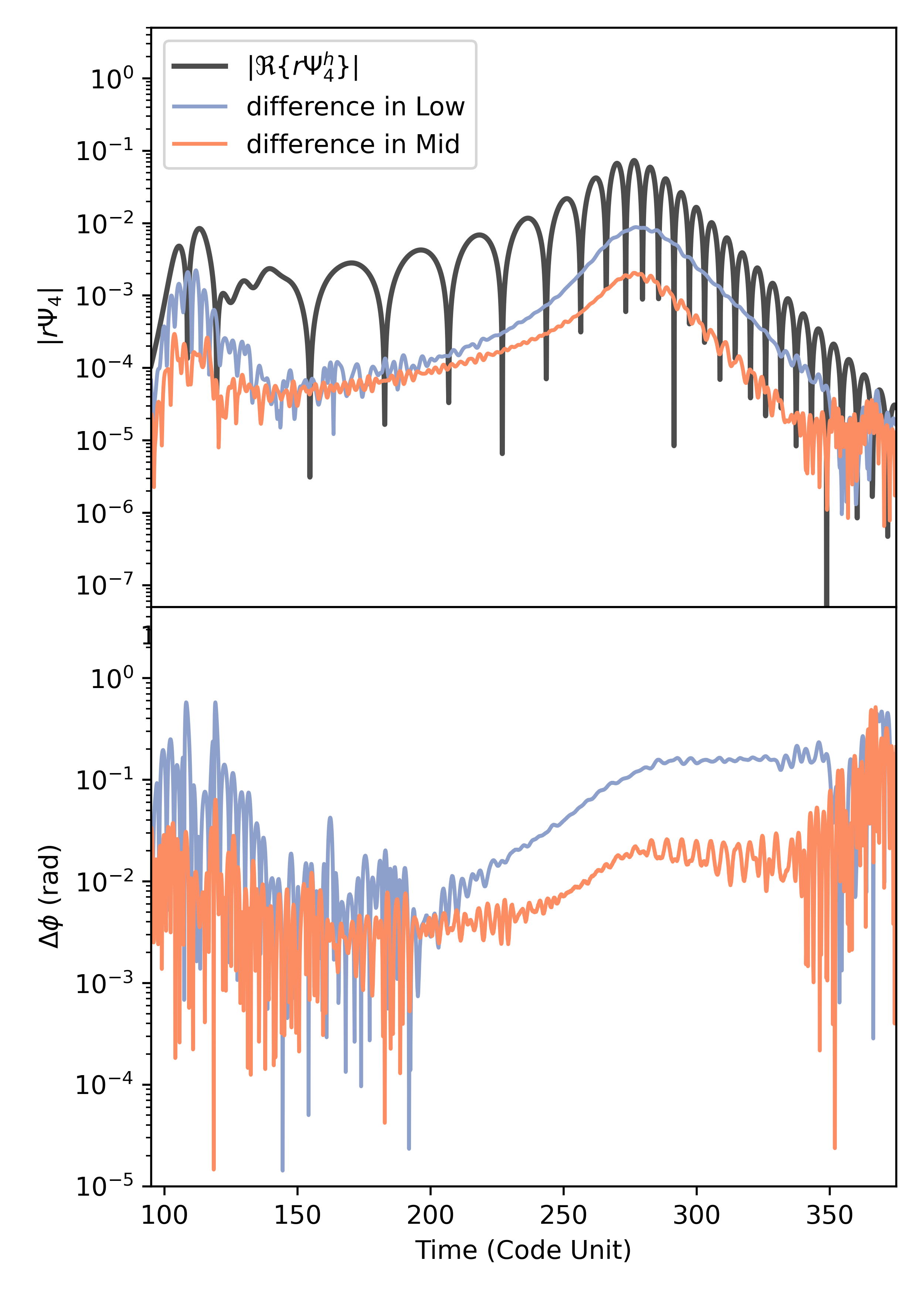}
  \caption{Waveform comparison between \AthenaK and \GRAthena. Like in Fig.~\ref{fig:conv}, the black curve shows the waveform amplitude at high resolution from \AthenaK. The purple and orange curves show the difference between the two codes for the low and mid resolution runs respectively. }
  \label{fig:comp++}
\end{figure}

In addition to self-convergence, we compare the waveforms with those from \GRAthena using the same initial data.
The available \GRAthena runs are done using the $L^2$ \AMR{} criteria, and with a K.O. dissipation of 0.02 (as oppose to 0.5 used here.)
The size (number of points along each dimension) of the \MeshBlock s are also significantly different, as smaller ($16^3$) \MeshBlock s are used to improve scaling on CPUs. 
Furthermore, the \GRAthena runs are done with Vertex-Centered scheme, as oppose to the Cell-Centered scheme for \AthenaK. 
We choose \GRAthena and \AthenaK simulations with matching resolution at the puncture, and compare the residual difference between the two codes at these two different resolutions (Low and Mid). 
Despite the aforementioned differences, we find that the difference between the codes still converges away with resolution (see Fig.~\ref{fig:comp++}).

\subsubsection{Accuracy and resolution requirement with AMR}
In this section we do a brief exploration of how the grid structure affects the accuracy of waveform. 
Starting with a low resolution run, to increase the accuracy of the simulation, one can either perform a global refinement, where the resolution throughout the \Mesh{} is uniformly increased. 
Alternatively, one can increase the maximum level of mesh-refinement, which will only increase the resolution around the punctures, leaving the wave-zone resolution untouched. 
The advantage of the later strategy is due to performance: while a global increase in resolution, say by a factor of 2, increases computational cost by a factor of 16, changing the maximum level of mesh-refinement by 1 only increases the computational cost by a factor of roughly 2. For the later strategy, the number of \MeshBlock s and thereby gridpoints only changes by around 10\% typically, so the increase in computational cost is primarily due to the CFL condition. The resulting grid structure yields matching resolution at the puncture but only half the resolution in the wave-zone.

\begin{figure}[t]
  \includegraphics[width=0.45\textwidth]{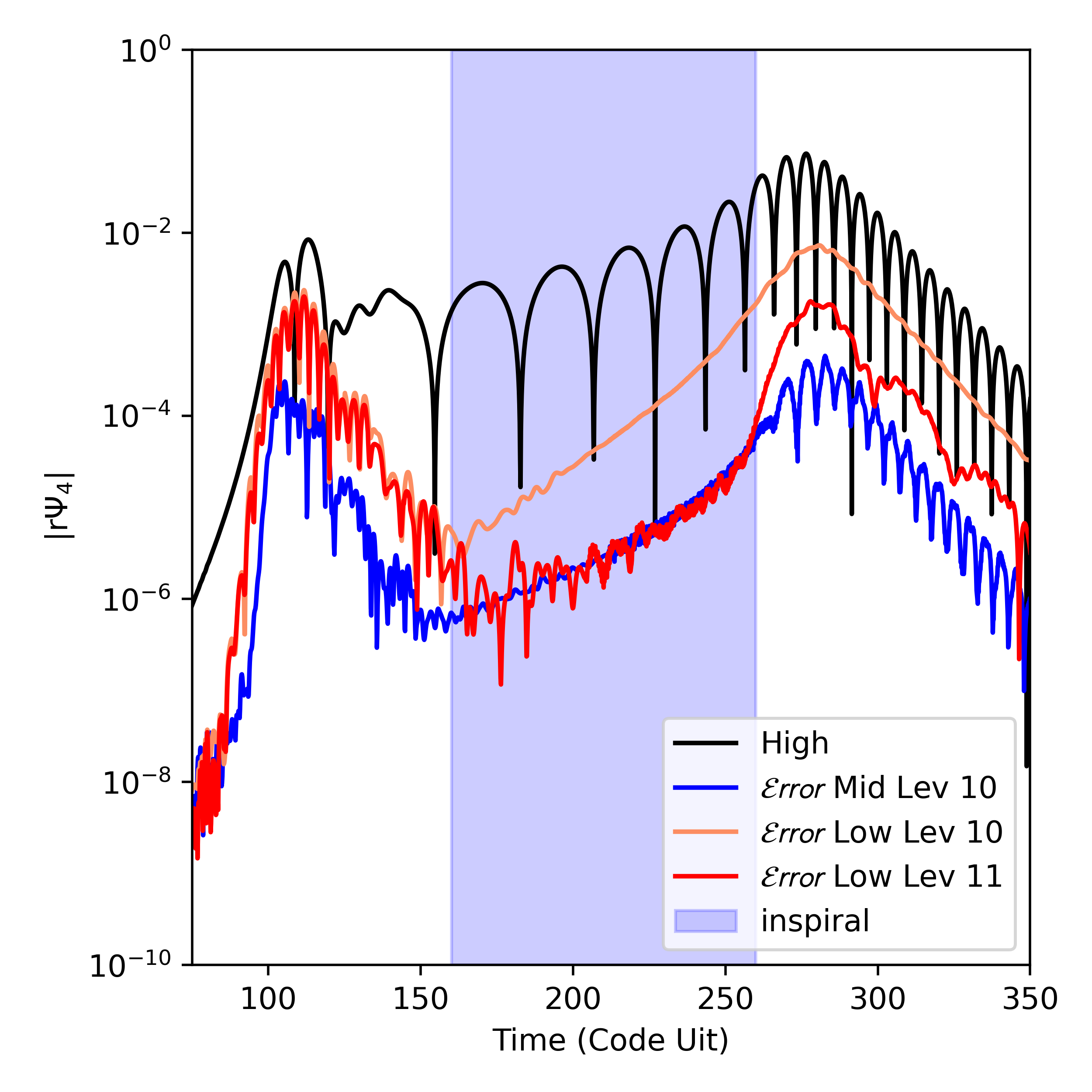}
  \caption{Comparison of error in gravitational waveform with different grid structures. The error is estimated by taking the difference between the waveform of a given run and that for the high resolution run. The Mid (orange) and Low Lev 10 (blue) shares the same mesh refinement structure, yet the Mid has twice the resolution within each \MeshBlock. Low Lev 11 (red) has the same root grid as Low Lev 10, yet one more level of mesh refinement, yielding the same resolution at the puncture as the Mid resolution run. We find that during the inspiral phase Low Lev 11 yields an error consistent to the Mid Lev 10 run, despite being roughly 8 times cheaper. }
  \label{fig:amr_test}
\end{figure}

In Fig.~\ref{fig:amr_test}, we plot the error in waveform for the calibration run with several different configurations. 
In addition to the Low and Mid resolution with 10 levels of mesh-refinement used in the convergence test, we now add an additional run using the same root grid as the Low resolution run but now with 11 levels of mesh-refinement. 
The error on the waveform is then estimated by the difference with the High resolution waveform. 
We find that, during the inspiral phase, as marked by the blue shaded region in Fig.~\ref{fig:amr_test}, the error between the new Low Lev11 run is practically the same with that for the Mid resolution run, despite being 8 times cheaper. However, this is no longer the case during the early nor late part of the waveform, corresponding to junk radiation and merger-ringdown phase respectively. This is primarily due to the presence of higher-frequency signals, which is not as well resolved in the wave zone. 

In typical binary black hole simulations, the inspiral phase is an order of magnitude longer than both the junk radiation and merger-ringdown phase, thereby increasing the level of refinement instead of a global refinement would save a tremendous amount of computing resources. The junk radiation phase does not need to be resolved, as it is typically cut out from the waveform (see, e.g., \cite{Pretto:2024dvx}). 
To resolve the merger-ringdown phase, one can simply increase the wave-zone resolution when the coordinate separation of the two punctures is below a certain threshold. 
We defer a detailed diagnostic to a future work. 

\section{Performance}\label{sec:scale}
As \AthenaK aims at solving exa-scale problems, we demonstrate the performance portability and scaling of the vacuum numerical relativity module on large GPU clusters. 
In this section, we first show performance portability across a wide range of hardware architectures. Furthermore, we show its scaling on Frontier and Perlmutter, which employs \texttt{AMD} and \texttt{NVIDIA} GPUs respectively. 
Through out this section, the performance is quantified by Zone-Cycle-Per-Second (ZCPS), or the number of cell updates per wall clock second, and all tests are done with $6^{\rm th}$ order spatial differencing and RK4 time integrator. 

\begin{table}
     \centering
     %\vspace{-.25in}
     \begin{tabular}{ccc} \\ \hline
     \textbf{Configuration} & 
     \textbf{ZCPS} & \textbf{1/Speed-up} \\
     \hline
     \texttt{NVIDIA} A100 & $1.36 \times 10^{7}$ & 1 \\
     \texttt{NVIDIA} Grace Hopper & $2.46 \times 10^{7}$ & 0.55 \\
     \texttt{NVIDIA} RTX 3070 Ti & $3.90 \times 10^{6}$ & 3.48 \\
     \texttt{AMD} MI250X (single die) & $5.08 \times 10^{6}$ & 2.68 \\
     \texttt{AMD} 7950x Zen 3 (single core) & $4.53 \times 10^{4}$ & 300.3 \\
     \texttt{Intel} Cascade Lake (single core)  & $1.71 \times 10^{4}$ & 793.65 \\
%     \texttt{Intel} Max Series GPU & -- & -- \\
     \texttt{Apple} M3 (single core; ARM) & $6.53 \times 10^{4}$ & 208.33 \\
     \hline
     \end{tabular}
     \caption{\label{tab:portability}Performance portability of \AthenaK Z4c module across a wide range of hardware architectures. Here, ZCPS is measured either per device or per CPU core. We also show the inverse speed up of each computing device compared against a mainstream \texttt{NVIDIA} A100 GPU. }
\end{table}

\subsection{Performance Portability}

Modern high performance computers employ a wide range of hardware architectures. 
Therefore, performance portability is arguably the most crucial feature for a code aiming at Exa-scale applications. 
To this end, we test the Z4c module in \AthenaK on a wide range of hardware, ranging from ARM and X86 CPUs to GPUs by different vendors. 
To measure the performance, we run the linear gravitational wave problem with $64^3$ \MeshBlock s. To make sure the GPUs are properly saturated, we load them with the largest number of \MeshBlock s that could fit into the Video Random-Access Memory (VRAM), whereas on CPUs we only use a single \MeshBlock{} as resource saturation is not a concern. 
We find that both \texttt{AMD} and \texttt{NVIDIA} GPUs offer a speedup of two orders of magnitude when compared to a modern CPU core. 
The performance numbers are summarized in Tab.~\ref{tab:portability}. 

\subsection{Weak scaling}
It is also important to demonstrate that the code can saturate resource on large computing clusters through weak scaling tests. 
For this test, we evolve a single black hole in puncture gauge with 10 levels of static mesh-refinement on Frontier.
To also show the dependence on the size of \MeshBlock, we perform the same test with three different \MeshBlock{} sizes, namely $32^3$, $64^3$, and $128^3$. 

We then double the resolution uniformly throughout the \Mesh{} and increase the resource by factors of 8 simultaneously, except for the largest run which uses 65,536 GPUs, where a factor of 8 increase on the resource exceeds the size of Frontier. 
For the last run, we only double the resolution along the x and y axis and increase the resource by a factor of 4 instead. 
We find that, with $64^3$ \MeshBlock s, scaling up from 4 to 65,536 GPUs retains an $80\%$ weak scaling efficiency, indicating that \AthenaK can efficiently run on exascale machines. The performance and scaling efficiency are shown in Tab.~\ref{tab:grav.scaling}. 

For the runs with $32^3$ \MeshBlock s, we are using twice the number of GPU  for the same problem because it cannot be fitted on to the same number of GPU due to the increased ratio of ghost zone to active zone. 
In general, we observe roughly a factor of 2 increase in performance when changing the size of the \MeshBlock{} from $32^3$ to $128^3$ on \texttt{AMD} MI250X. 

\begin{figure}[t]
  \includegraphics[width=0.45\textwidth]{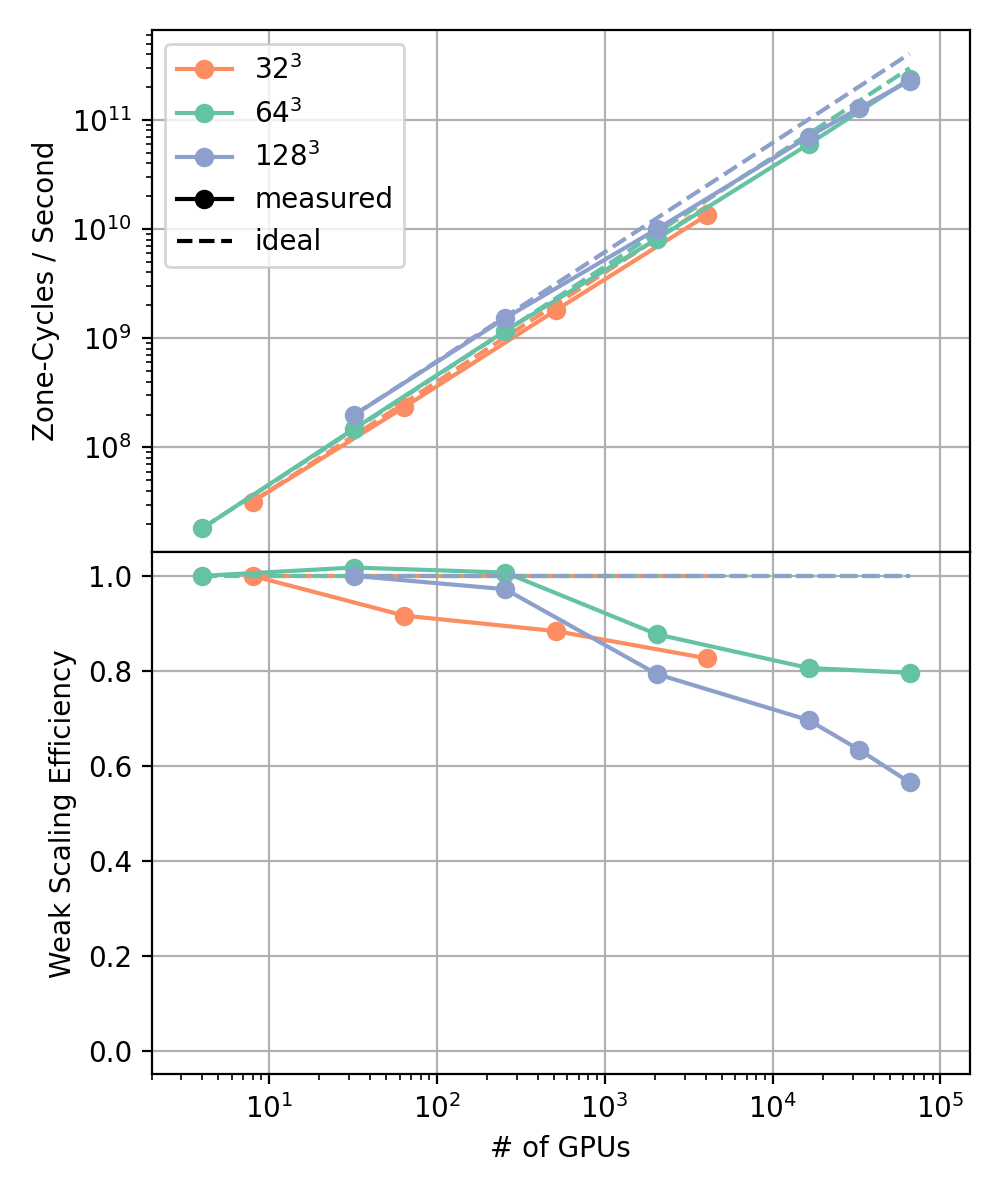}
  \caption{Weak scaling performance (top) and efficiency (bottom) for evolving a single black hole on Frontier with varying \MeshBlock{} sizes.} 
  \label{fig:weak_scaling}
\end{figure}

\subsection{Strong scaling}

To enable rapid scan of the binary black hole parameter space, excellent strong scaling is required to shrink the wall clock for each simulation. 
For this test, we perform the same test with 4,544 $64^3$ \MeshBlock s. 
The size of this test problem is tuned to match those for high resolution binary black hole runs with high mass ratio or high spin. 
The resulting resolution at the puncture is $\sim$ 0.008 M, same as the high resolution binary black hole calibration run discussed in section~\ref{sec:bbh}. 

We perform the test on both Frontier and Perlmutter. 
Different from CPU computing, the size of VRAM is the limiting factor for how many GPUs are needed to evolve a given problem. 
Frontier utilizes AMD MI250X, each with two dies (two logical GPUs) and a total of 128 GB VRAM, whereas Perlmutter uses the 40 GB variant of NVIDIA A100. 
Due to the limitation of VRAM size per device, the problem can be fitted onto 32 logical GPUs on Frontier, whereas on Perlmutter we uses 64 GPUs at the smallest.

\begin{table*}
     \centering
     \vspace{-.25in}
     \begin{tabular}{rcccccc} \\ \hline &
     \multicolumn{2}{c}{\textbf{Weak Scaling (Frontier)}} &
     \multicolumn{2}{c}{\textbf{Strong Scaling (Frontier)}} &
     \multicolumn{2}{c}{\textbf{Strong Scaling (Perlmutter})} \\ \hline
     \textbf{\# GPU} & 
     \textbf{ZCPS} & \textbf{Efficiency} &
     \textbf{ZCPS} & \textbf{Efficiency} &
     \textbf{ZCPS} & \textbf{Efficiency} \\
     \hline
     4 & $1.81 \times 10^7$ & 1.00 & $-$ & $-$ & $-$ & $-$\\
     32 & $1.48 \times 10^8$ & 1.02 & $1.48 \times 10^8$ & 1.00 & $-$ & $-$ \\
     64 & $-$ & $-$ & $3.26 \times 10^8$ & 1.10 & $6.04 \times 10^8$ & 1.00 \\
     128 & $-$ & $-$ & $7.10 \times 10^8$ & 1.20 & $1.08 \times 10^9$ & 0.89\\
     256 & $1.17 \times 10^9$ & 1.01 & $1.30 \times 10^9$ & 1.10 & $1.89 \times 10^9$ & 0.78\\
     512 & $-$ & $-$ & $2.39 \times 10^9$ & 1.01 & $5.45 \times 10^9$ & 1.13\\
     1024 & $-$ & $-$ & $3.96 \times 10^9$ & 0.84 & $9.51 \times 10^9$ & 0.98\\
     2048 & $8.14 \times 10^9$ & 0.88 & $5.76 \times 10^9$ & 0.61 & $1.49 \times 10^{10}$ & 0.77\\
     16384 & $5.99 \times 10^{10}$ & 0.81 & $-$ & $-$ & $-$ & $-$\\
     65536 & $2.36 \times 10^{11}$ & 0.80 & $-$ & $-$ & $-$ & $-$\\
     \hline
     \end{tabular}
     \caption{\label{tab:grav.scaling}Strong and weak scaling data with $64^3$ \MeshBlock s. The scaling efficiency is computed by the run with the smallest number of GPUs for all cases. }
\end{table*}

\begin{figure}[t]
  \includegraphics[width=0.45\textwidth]{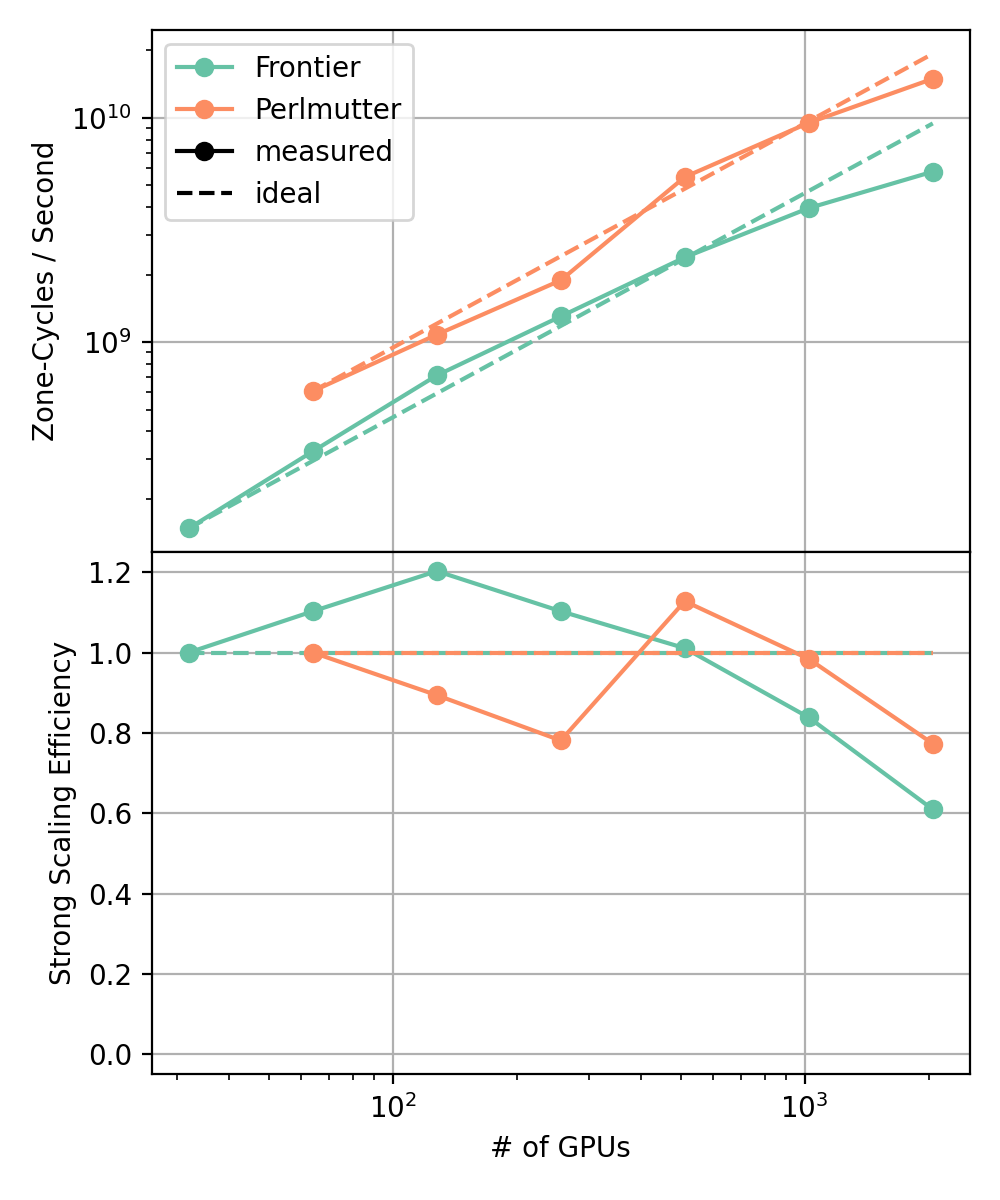}
  \caption{Strong scaling performance (top) and efficiency (bottom) for evolving a single black hole with $64^3$ \MeshBlock{} on both Frontier and Perlmutter.}
  \label{fig:strong_scaling}
\end{figure}

The performance numbers are shown in Fig.~\ref{fig:strong_scaling} and Tab.~\ref{tab:grav.scaling}.
We find that \AthenaK still maintain 84\% and 77\% efficiency on Frontier and Perlmutter respectively when scaling up the computational resource by a factor of 32, though the number drops significantly to 61\% on Frontier when we increase the resource further by a factor of 64. 
This is not unexpected on GPUs, as the minimum amount of workload to saturate a GPU is much larger than that for CPUs. 
We also report a factor of two increase in performance per logical device for \AthenaK when running on \texttt{NVIDIA} A100 compared with \texttt{AMD} MI250X (single die). 

\section{Discussion and Conclusion} \label{sec:conclude}
In this paper, we introduced and validated the numerical relativity module within \AthenaK, a re-implementation of \GRAthena optimized for Exa-Scale applications using the \texttt{kokkos} library. 
Our focus was on solving the vacuum Einstein equations in the Z4c formulation with Oct-Tree Adaptive Mesh Refinement (AMR), addressing the need for high-accuracy gravitational waveforms with large parameter space coverage, which is crucial for next-generation gravitational wave detectors.

We demonstrated \AthenaK's accuracy and scalability through convergence of linear gravitational wave, accuracy of spinning puncture evolution, and convergence of gravitational waveforms from equal-mass black hole binaries, cross-code validation with \BAM and \GRAthena. These results confirm \AthenaK's reliability for precision waveform modeling and its ability to produce consistent, accurate results.

We further demonstrate performance portability across all major high performance computing architectures. Particularly, scaling tests on Frontier and Perlmutter, using AMD and NVIDIA GPUs, respectively, demonstrated AthenaK's excellent performance on Exa-Scale platforms. The code achieved 80\% weak scaling efficiency up to 65,536 GPUs on Frontier and approximately 80\% strong scaling efficiency when increasing the number of GPUs by a factor of 32 on both Frontier and Perlmutter. 
These results underscore AthenaK's capacity to leverage the computational power of emerging Exa-scale computers, making it a powerful tool for large-scale numerical relativity simulations.

\acknowledgments

We thank the many contributors to the {\tt Athena++} code project. 
We thank Patrick Mullen and Frans Pretorius for valuable conversations, and Matthew Cawood for performance numbers on NVIDIA's Grace Hopper Chip. 
HZ thank Frans Pretorius and Nils Vu for providing comments on the draft. 
% The simulations presented in this article were performed partly on Perlmutter
DR was supported by funding from the U.S. Department of Energy, Office of
Science, Division of Nuclear Physics under Award Number(s) DE-SC0021177 and DE-SC0024388,
by NASA under award No. 80NSSC21K1720, and by the National Science Foundation under Grants
No. PHY-2011725, PHY-2020275, PHY-2116686, and AST-2108467.
JS was supported by a subcontract from the Texas Advanced Computing Center from National Science Foundation grant 2139536, and by the Eric and Wendy Schmidt Funds for Strategic Innovation. 
FZ, SB and BD acknowledge funding from the EU H2020 programme under ERC Starting Grant, no.~BinGraSp-714626
SB and BD acknowledge funding from the EU Horizon programme under ERC Consolidator Grant, no.~InspiReM-101043372.

Simulations were performed on OLCF's Frontier, NERSC's Perlmutter, Pennsylvania State University's Institute for Computational and Data Sciences's Roar Collab supercomputer, and on Princeton University's Della and Stellar supercomputers.
This research used resources of the National Energy Research Scientific
Computing Center, a DOE Office of Science User Facility supported by the Office of Science of the U.S.~Department of Energy under Contract No.~DE-AC02-05CH11231.  
This research used resources of the Oak Ridge Leadership Computing Facility at the Oak Ridge National Laboratory, which is supported by the Office of Science of the U.S. Department of Energy under Contract No. DE-AC05-00OR22725.
The simulations presented in this article were partly performed on computational resources managed and supported by Princeton Research Computing, a consortium of groups including the Princeton Institute for Computational Science and Engineering (PICSciE) and the Office of Information Technology's High Performance Computing Center and Visualization Laboratory at Princeton University.

\bibliographystyle{apj} % if there are any compatibility issues when compiling on Overleaf
\bibliographystyle{plainnat}
\bibliographystyle{unsrt} % numbered references, listed in order of appearance
\bibliography{ref}
\end{CJK*}
\end{document}